# Thermal mirror buckling in freestanding graphene locally controlled by scanning tunneling microscopy

M. Neek-Amal[1,#], P. Xu[2,3], J. K. Schoelz[2], M. L. Ackerman[2], S. D. Barber[2], P. M. Thibado[2]*, A. Sadeghi[4] and F. M. Peeters[1]*

[1]*Departement Fysica, Universiteit Antwerpen, Groenenborgerlaan 171, B-2020 Antwerpen, Belgium.*

[#]*Department of Physics, Shahid Rajaee Teacher Training University, Lavizan, Tehran 16788, Iran*

[2]*Department of Physics, University of Arkansas, Fayetteville, Arkansas 72701, USA.*

[3]*Department of Physics, University of Maryland, College Park, MD 20742, USA*

[4]*Departement Physik, Universität Basel, Klingelbergstrasse 82, CH-4056 Basel, Switzerland.*

*e-mail: thibado@uark.edu; francois.peeters@uantwerpen.be

**Abstract**

Knowledge of and control over the curvature of ripples in freestanding graphene is desirable for fabricating and designing flexible electronic devices, and recent progress in these pursuits has been achieved using several advanced techniques such as scanning tunneling microscopy (STM). The electrostatic forces induced through a bias voltage (or a gate voltage) have been used to manipulate the interaction of freestanding graphene with a STM tip (substrate). Such forces can cause large movements and sudden changes in curvature through mirror buckling. Here we explore an alternative mechanism, thermal load, to control the curvature of graphene. We demonstrate thermal mirror buckling of graphene by STM experiments and large-scale molecular dynamics simulations. The negative thermal expansion coefficient of graphene is an essential ingredient needed in explaining the observed effects. This new control mechanism represents a fundamental advance in understanding the influence of temperature gradients on the dynamics of freestanding graphene, as well as future applications with electro-thermal-mechanical nanodevices.

**Main text**

Graphene has the required physical properties to provide the foundation for a technological revolution. From high-performance flexible electronics to tabletop experiments in relativistic quantum mechanics, the range of possibilities seems unlimited[1]. Interestingly, the mere existence of this new material is perhaps its most astounding feature[2]. Despite the impossibility of long-range stability in any two-dimensional (2D) crystal, which was well-established theoretically by Peierls, Landau, and the Mermin-Wagner theorem[3], the anharmonic coupling between bending and stretching phonons[4, 5] stabilizes the 2D crystal, thereby establishing that deviations from planarity are essential to the stability of isolated graphene. Subsequent experiments provided evidence that random nanoscale roughening does exist and is manifested as ripples approximately 0.5 nm high and 5-10 nm wide[6-8]. In fact, when pristine suspended graphene is viewed via transmission electron microscopy[9] or scanning tunneling microscopy (STM)[10], its topography resembles a network of adjacent hemispherical surfaces opening alternately upward and downward. Yet, this natural intrinsic roughening is not the only allowable configuration; it is possible, in effect, to rearrange the ripples to achieve lattice distortions of a desired shape, size, or periodicity through strain engineering[11-14].

In fact, many important nanoelectromechanical (NEM) graphene device concepts have been recently developed based on the electro-mechanical properties of graphene. For example, Chen et al. demonstrated that graphene can be used as a nanomechanical resonator with an electrical readout that varied with temperature or added mass[15]. Shortly after Park et al. demonstrated a graphene mechanical actuator that responded to chemical changes, among other things[16]. At this same time, Mashoff et al. built a NEM graphene membrane and demonstrated bi-stability control using STM[13]. This led to Lindahl et al., exploiting the snap-through instability of pre-buckled suspended graphene in a NEM device and they measured its operating voltage characteristics[17]. The most recent development was by Eder et al., where they placed an STM tip on either side of a freestanding graphene film and demonstrated tunable membrane deformations using electrostatic control[18]. As far as thermally-induced mechanical movement in real devices is concerned, thermal actuation of a microelectromechanical (MEM) device (i.e., silicon-on-insulator technology) was first introduced because thermal loads provide a significantly larger force as compared to electrostatic actuation[19]. Heating under the STM tip has been studied experimentally and a giant enhancement in electronic tunneling at higher energies due to an intrinsic phonon-mediated inelastic channel was found to be responsible for an unexpected gap-like feature in the graphene tunneling spectrum[20]. Progress with theoretical estimates for the heating have also been made by studying inelastic currents through nanoscale molecules sandwiched between gold electrodes, for example[21, 22].

In this article, we control the local height and curvature of freestanding graphene by varying the STM tunneling current. The movement of the graphene membrane can be tuned to vary smoothly or in step-like jumps attributed either to electrostatically-induced mirror buckling or to tunneling current-induced (i.e., thermally-induced) mirror buckling. The observed thermal buckling is explained by both elasticity theory and large-scale molecular dynamics (MD) simulation. Standing on the body of work and combined with our discovery of negative thermal buckling we propose an electro-thermo-mechanical (ETM) device.

**STM imaging and Scanning Tunneling Spectroscopy (STS) thermal buckling**

Given that graphene is all surface, STM is becoming the tool of choice to manipulate and map suspended graphene. As the biased STM tip approaches a naturally rippled freestanding graphene surface, depicted in Fig. 1a, the two are drawn together electrostatically. A typical constant-current, filled-state STM image, measuring 6 nm × 6 nm and acquired using a tip bias voltage of 0.1 V and a tunneling current of 1.0 nA, is displayed in Fig. 1b with a 4 nm black-to-orange-to-white height scale. (See Methods for further STM and sample details.) The characteristic honeycomb structure, though distorted by sample movements, is visible throughout the image, and the overall topography features a wide ridge running diagonally from the bottom left corner to the top right corner. Note that these are difficult images to obtain because unsupported graphene is typically very floppy by STM standards[10]. An important effect caused by the STM tip under certain conditions is the local heating of the sample by passing current through it[23]. In order to demonstrate this with freestanding graphene, the tunneling current was ramped from 0.01 nA to 20 nA at a constant tip bias of 20 mV in a feedback-on configuration, and the result is shown in Fig. 1c (Our control sample result was acquired using graphene on copper which is also shown in Fig. 1c). This measurement reveals that the tip height drops an astonishing 20 nm over the first 7 nA, beyond which point it fluctuates strongly around this minimum height. One must simultaneously record the actual tunneling current throughout this measurement to ensure the feedback circuit is able to maintain the specified setpoint tunneling currents. That data are shown in the inset of Fig. 1c, confirming this condition was achieved. Therefore, the drop in height occurred because graphene is physically pulling away from the STM tip as the tunneling current is increased. It cannot be due to the decreasing tunneling gap, which would only be on the order of angstroms[24]. It also cannot be due to changes in the electrostatic force, because reducing the tunneling gap would increase the attractive force, resulting in a height increase rather than a decrease. Instead, what is happening as we increase the tunneling current is that graphene under the tip is being locally heated (i.e., Joule heating) by the additional tunneling current. Given that graphene has a negative thermal expansion coefficient, the film contracts away from the tip as the temperature under it is increased. It is possible to estimate the temperature of the graphene directly under the STM tip. Using a value of $(-10^{-5})$ K$^{-1}$ for the thermal expansion coefficient[25], our height contraction of 20 nm, and assuming a distance of 3.5 μm to the copper support (which is assumed to remain at room temperature), an increase of 10–200 K is estimated (For more theoretical details see Supplementary Note 1).

A second important effect caused by the STM tip, already mentioned, is the electrostatic attractive force due to the biased tip being adjacent to the grounded sample, which allows us to pull the freestanding graphene. Two typical constant-current tip displacement ($Z$) data sets as a function of bias voltage ($V$) for a range of tunneling current setpoints ($I$) are displayed in Fig. 2a. One data set is for positive bias voltage sweeps from 0.1 V to 3.3 V, while the other is for negative voltage sweeps from −0.1 V to −3.1 V. All the positive voltage sweeps were taken at the same location on the sample, while the negative bias data was collected at a new location more than 10 μm away. The 0.1 nA and 10 nA curves presented are averages over ten consecutive measurements, but the 2 nA and 4 nA trials shown, where *a sudden permanent jump in height occurred*, are single runs. Note, data sets were actually collected in smaller current step sizes (i.e., 0.1 nA, always increasing), but for clarity only the three characteristic types are shown [All the Z(V) data are shown in Supplementary Fig. 1]. The low-current curves (red, 0.1 nA) are characterized by a noticeable increase in tip height (approximately 30–35 nm) as the tip bias is ramped. They are also reversible and repeatable with reasonable regularity. Next, the 4.0 nA curve (black) shows two small jumps around 0.6 V and 1.1 V, a long plateau, and then displays a large jump of about 35 nm at 3 V before falling slightly. The large jump in height was permanent, as demonstrated in a moment. For the negative bias voltages, a permanent jump occurred during the 2.0 nA trial at −1 V, the difference presumably due to different initial conditions at the new sample location (please see Supplementary Fig. 2 for a plot of the height and current measured during the negative voltage sweep 2.0 nA data set). Finally, the high-current curves (blue, 10 nA) show a total tip height change of only 3–4 nm over the entire bias range. The high-current curves are also reversible and repeatable. They are displaced at the top of the graph because the sample had previously shifted to that height and remained there (Note, a much larger number of data sets showing the thermal mirror buckling event were also acquired and a summary plot showing the bias voltage for which the buckling event occurred versus the setpoint current for which the event occurred is shown in Supplementary Fig. 3).

During the height-voltage sweeps, it is important to simultaneously record the tunneling current to ensure the tip is tracking the movement of the sample. The measured tunneling current for the 4.0 nA data set is constant for most of the voltage range, as shown in Fig. 2b. However, a very high, narrow peak is observed at ~3.0 V. This is due to the sample approaching the tip too quickly for the feedback circuit to keep the current constant[26]. However, we can confirm that the sample does not "crash" into the tip because our system's saturation current (50 nA) is not reached. In addition, it is important to notice that shortly after the surge in current, it returns to the setpoint, and stable tunneling is once again achieved for the remaining voltage sweep (please see Supplementary Fig. 4 for a reduced voltage range plot showing individual current data points near the 3 volt spike in current). Topographic data was simultaneously recorded during the time in which the 4.0 nA $Z(V)$ curve was collected, and a line profile extracted from that height data are shown in Fig. 2c. It represents 400 s in time and reveals that a permanent increase of ~80 nm occurred immediately after the 4 nA measurement was taken. Based on this permanent jump in the height of the sample, the $Z(V)$ curve collected for higher current is offset accordingly, which places it above the $I$ = 4.0 nA data set (a similar offset was observed for negative bias voltages) [Note, it is sometimes possible to thermal mirror buckling the graphene downward using an even higher current as shown in Supplementary Fig. 5].

The large, permanent, and sudden jump in the height of the graphene film has been observed before and is classified as mirror buckling[17, 27]. An illustration of the new surface configuration is shown in Fig. 2d. Mirror buckling refers to the sudden reversal of a dimple in a thin film, causing it to go from concave to convex or vice versa. Typically, an electrostatic force of a certain magnitude is sufficient for causing the mirror buckling in graphene. The fascinating aspect for each of our data sets is that there is a transition from a *smooth trajectory* to a *step-like trajectory* occurring with increasing current, that is, when heating up the sample. Observation of the mirror buckling effect still requires the aid of an electrostatic force, but it is not solely responsible for the emergence of the large jumps (e.g., the large jumps never happen at lower currents, so higher currents are also required). We believe this unexpected behavior can be understood as thermal-induced mirror buckling when taking into account the role of the negative thermal expansion coefficient of graphene, and we now discuss our theoretical model for such events.

**MD simulations of thermal buckling**

Atomistic simulations can provide a deep understanding of the observed phenomena (Please see Supplementary Fig. 6 and Supplementary Note 2 for our continuum elasticity theory approach). Therefore, in this section the dynamic aspects of freestanding graphene ripples, interacting with the STM tip, are studied by performing MD simulations. For our computer model, we created a circular graphene sheet with a diameter of 0.18 μm containing 1.1 million carbon atoms, depicted in Fig. 3a. The nearly micron size of our sample was necessary because many different length scales are present in this problem, and therefore it was critical for capturing the important collective behavior. Due to the vastness of the system, it was also necessary to narrow the scope of the simulations. It would be impossible to simulate the entire experiment because the bias voltage is swept over two decades, the current is swept over three decades, and the experiments run for hours. We focus our attention on what is new and interesting in the MD experiments, which is *the role of temperature on the mirror buckling process*. To do this, in brief, we prepared an already convex buckled graphene state by placing it under the influence of a 3 V bias voltage, as shown in Fig. 2d. A central region of the sheet ($r < 10$ nm) was kept at a temperature $T_c$, while the outer boundary was held at 300 K. An example equilibrium temperature profile created with the central temperature set to $T_c = 500$ K is shown in Fig. 3b. Notice the temperature profile is nonlinear. See Methods for further details on the computational model.

The graphene height distribution for our starting configuration, having applied a bias voltage of 3 V, is represented in Fig. 4a. The central region is attracted to the tip and has buckled toward it by about 11 nm. Notice the oscillating height when moving in a circular pattern around the outer boundary with an amplitude of about 3 nm and a wavelength of about 30 nm. These features play a subtle but important role in the collective behavior of the system. To demonstrate thermal buckling with our MD simulations, we changed the bias voltage to 0.22 V, instantaneously increased the central temperature to 500 K, and observed the system's evolution in time. The graphene height distribution 50 ps after changing the temperature is plotted in Fig. 4b. Since the system was initially convex, the top of the graphene, after additional heating, is expected to buckle downward, making it concave. This feature can, in fact, be seen near the center of the plot as a depression in the height. Vertical line profiles were extracted from both density plots at $x = 0$ nm and are shown in Fig. 4c. The upper height profile (red curve) is before the temperature increased, while the lower (blue curve) is 50 ps after. The central features are further magnified in Fig. 4d for clarity. What is important is that the height of the central region reversed by about 3 nm after changing the temperature to 500 K, while at the edges the line profile did not change its height. In fact, all that what has happened is that the curvature near the center has flipped (i.e., from convex to concave) and has left 30° sharp bends in the mirror-buckled line profile.

We can confirm that the central height change is due to heating by comparing these results with a new simulation where the bias is set to 0.22 V but $T_c$ remains 300 K, as shown in Fig. 5a. The top two curves (red) show the height of the central atom (solid line) and the average height of the whole central region (dashed line) as a function of time when the temperature of the central region is left at 300 K. The central atom undergoes noticeable fluctuations throughout, and the average height actually increases by about 0.5 nm. On the other hand, when $T_c$ = 500 K the lower curves (green) demonstrate a dramatic drop in average height, as well as much larger height fluctuations of the central atom. The downward movement is due to the negative thermal expansion coefficient of graphene, and is in agreement with the experimental results shown in Fig. 1d.

The above simulation addressed what happens when the STM tip is over a locally convex region of graphene (see Fig. 5a, lower illustration). In that case, the electrostatic force and the thermal load work in opposite directions and compete against one another. A more dramatic effect is observed when placing the tip over a concave region, as shown in Fig. 5b (lower illustration, shown with tip underneath the graphene only for consistency in the direction of the height change). The MD simulation starts with the same initial configuration as before. However, now increasing $T_c$ to 500 K results in the sample contracting toward the tip (due to its new curvature), making the height decrease more rapidly. In fact, the overall displacement was so large that the entire MD sample flipped over within the same 50 ps time period (i.e., it reached negative heights). We can understand this dramatic behavior by realizing that the electrostatic force and thermal load now work in the same direction. Moreover, this matches well the effect observed for the middle currents in Fig. 2a. Again, without the temperature change (i.e., if the central temperature is left at 300 K), the system only drifts toward the tip by about 2 nm as shown in the upper curves of Fig. 5b.

**Discussion**

When heating a convex or concave region of graphene, the area will mirror buckle. For the convex starting case, it buckles away from the STM tip, while for the concave starting case, it buckles toward the STM tip. In this way, applying heat to the sample results in either an effective attractive force or a repulsive force. This is not the case with the applied bias, where both positive and negative voltages result in only an attractive force being applied to the sample. With this unique feature in mind, we can now fully understand the sudden jumps in the sample. For low tunneling current, the sample height simply follows the electrostatic force up and down. But by raising the tunneling current, graphene will first buckle away from the STM tip (e.g., see Fig. 1c), such that when we perform the voltage sweep, the system is primed for a major mirror buckling event. This appears to be the mechanism behind the jumps observed in Fig. 2a for tunneling currents of 4 nA and 2 nA, placing them in good agreement with our MD simulations.

One primary difference between the MD and experimental STS results is the size of the jumps. In the STS experiment, the jump is about an order of magnitude larger than that of our MD simulations, but this is reasonable because the size of the MD sample is about an order of magnitude smaller (i.e., the ratio of sample size 0.2 μm/7.5 μm is proportional to the ratio of the largest jump). Another important difference is the time scale for the simulation versus the experiment. For the STS measurement, the time is on the order of seconds. We believe that, although the simulation time is many orders of magnitude smaller than the real time, the mirror buckling effect itself simply happens too quickly for the STM to track it instantaneously. Plus, we increase the current slowly in the STM, whereas the MD temperature change happens instantaneously. Furthermore, when we remove the bias voltage in the MD simulation, the initially buckled shape of the system (from applying 3 V) is stable, at least up to our maximum time of 200 ps, and there is no significant change in the height of the central atom. We speculate that this indicates that the theoretically calculated effect can also be found at larger time scales for larger bumps.

In a broad picture, we can conclude that as the STM voltage is increased, some of the suspended graphene ripples reverse their orientation and provide a mechanism for larger perpendicular displacement[28]. The larger the bias voltage, the greater the ability for reversing larger bumps. We can also conclude that as the current is increased, graphene is heated locally and contracts. This contraction is not to be understood in terms of a decreasing bond length, but rather an increasing amplitude of flexural phonon modes which causes an effective in-plane contraction[25]. The contraction increases the elastic energy, thereby making the system more unstable, such that when more voltage is applied at higher currents, the system can suddenly jump to form a larger stable structure. Given that the electrostatic force is proportional to $V^2$, it remains an attractive force upon flipping the bias. On the other hand, the tunneling current heats the freestanding graphene and can yield either a repulsive or attractive force, depending on the local curvature of the graphene sample. All totaled, we have developed a unique non-contact capability with STM to apply both attractive and repulsive forces through the vacuum. These forces result in controlled mechanical movement forming a foundation for a new type of ETM device, which also offers an ultra-sensitive mechanism for dynamic force sensing and related fundamental investigations.

Finally, thermal actuation is nicely illustrated by the larger height change in our MD results shown earlier. Equally important, however, thermal actuation also provides a new control mechanism which can push or pull on the surface, and makes possible dual electro-thermal control. Also, given graphene's negative thermal expansion coefficient, this thermal control is unique and opposite from other materials leading to new opportunities[25]. An ETM device is schematically shown in a six part illustration in Fig. 5c-h. Each part shows two input leads and one output. The inputs can be controlled using both temperature and voltage. For the starting configuration shown in Fig. 5c, the graphene membrane is shown in green and is connecting input 0 to the output. When heat is added to input 0 as shown in Fig. 5d the membrane is heated and thermally buckles to the input 1 lead as shown in Fig. 5e. Input 0 can go back to its original temperature to create the final state shown in Fig. 5h. Alternatively, a higher bias voltage can be added to input 1 as shown in Fig. 5f. This will pull the graphene membrane to the input 1 lead as shown in Fig. 5g. Input 1 can then go back to its original voltage to again create the final state shown in Fig. 5h. The temperature, $T_S$ required to switch the membrane position as well as the voltage, $V_S$ required to switch the membrane position are estimated in the supplement as a function of the size of the graphene membrane. Even though the MD simulations used here explain many different aspects of the observed phenomena, the system is richly complicated and demands more theoretical studies.

In summary, the effect of tunneling current on the ripples of freestanding graphene in STM measurements was investigated both experimentally and theoretically. A systematic series of STM experiments demonstrated that heating the sample significantly changes its response when pulled via an electrostatic force. Rather than a smooth increase the height under the tip as expected for a simple elastic sheet, we observed step-like jumps and plateaus, which are the result of a combination of mirror buckling and negative thermal expansion coefficient of graphene. This behavior was simulated in detail through MD performed on an exceptionally large sample. The MD results showed that the static ripples in graphene are very sensitive to the local temperature, and that a non-linear temperature profile can lead to sudden jumps in the height of the graphene film as observed experimentally. This collection of results provides unprecedented insights into the role of the thermal load in STM on freestanding graphene and complements previous work focusing on electrostatically induced buckling. Additionally, the extreme thermal sensitivity of the freestanding graphene membrane will have repercussions on electronic modifications of this still relatively unexplored system.

**Methods**

**STM experiments.** An Omicron ultrahigh-vacuum (base pressure is $10^{-10}$ mbar), low-temperature model STM, operated at room temperature, was used to obtain constant-current STM images of freestanding graphene, as well as feedback-on measurements of tip height at a single point as a function of either bias voltage or setpoint current. To perform such a measurement, a topography scan is already in progress (typically 0.1 nm by 0.1 nm), a point in the image is selected, and the imaging scanner is moved to and paused at that location long enough to sweep the voltage or current and measure the height before returning to its previous position to continue the topographic scan. The feedback loop controlling the vertical motion of the STM tip remains operational all the while. Assuming the sample is stationary, this process indirectly probes its density of states[29, 30]. A second interaction is also taking place, though, in which the tip bias induces an image charge in the grounded sample, resulting in an electrostatic force that increases with the bias and attracts the sample toward the STM tip. Freestanding graphene is flexible and responds to this force, so it cannot be assumed to be stationary.

The graphene was grown using chemical vapor deposition[31], then transferred by the commercial provider onto a 2000-mesh, ultrafine copper grid, consisting of a lattice of square holes 7.5 μm wide and bar supports 5 μm wide. At the STM facility, this grid was mounted on a flat tantalum sample plate using silver paint and transferred through a load-lock into the STM chamber, where it was electrically grounded for all experiments. In this system, the tip points upward at the downward-facing sample surface. Data was acquired using tips manufactured in-house by electrochemical etching of polycrystalline tungsten wire, using a custom double-lamella setup with an automatic gravity-switch cutoff. After etching, the tips were gently rinsed with distilled water, briefly dipped in a concentrated hydrofluoric acid solution to remove surface oxides, and loaded into the STM chamber.

**MD simulations.** The circular sheet of graphene in the computational model was divided into four different regions. We used the AIREBO potential which is particularly well suited for simulating properties of hydrocarbon systems[32]. First, there is a central part with $r < 10$ nm which is directly below the STM tip. The whole central region is kept at a constant temperature $T_c$, which was altered to simulate the different tunneling current setpoints. Second, the boundary region with width 0.2 nm (89.8 nm $< r <$ 90 nm) is held fixed and subjected to a very small shear strain (0.1°) in space. It defines the zero height position for all the simulations. Third, the outer region (88 nm $< r <$ 89.8 nm) is kept at a constant temperature of 300 K during the simulations. Fourth, there is the in-between area with all the remaining atoms, where temperature is calculated and its distribution is governed by the temperature of the central and outer divisions.

In order to include the effect of the STM tip bias voltage $V$, we modeled the tip-sample system as a capacitor whose capacitance $C$ is determined by the geometry of the tip-apex and the tip-sample separation. To model the role of the tip-apex, we distributed the charge $q = CV$ according to a Gaussian distribution of width $\sigma = 10$ nm (i.e., the size of the central region) over the atoms of the graphene layer, and we assumed that the local electric field $E$ is uniform over this region. Both $C$ and $E$ are determined by solving the boundary value electrostatic problem using a finite difference method[33]. In this way, an electric force $F_i = q_i E$ is applied normal to the surface at each atom $i$ during the MD simulation. Since both the electric field and the atomic charges $q_i$ are proportional to the bias voltage, the electric force is proportional to $V^2$.

**Acknowledgements**

Financial support was provided, in part, by the Office of Naval Research under grant N00014-10-1-0181, the National Science Foundation under grant DMR-0855358, the EU-Marie Curie IIF postdoc Fellowship/299855 (for M.N.A.), the ESF-EuroGRAPHENE project CONGRAN, the Flemish Science Foundation (FWO-Vl), and the Methusalem Foundation of the Flemish Government.

**Figure captions**

**Figure 1 | STM imaging and analysis. a**, Illustration of a biased STM tip near suspended graphene with intrinsic roughness. **b**, Constant-current, filled-state STM image of pristine freestanding graphene, measuring 6 nm × 6 nm and taken with $V = 0.1$ V and $I = 1.0$ nA. Total height range represented is 4 nm. **c**, Constant tip bias (20 mV), feedback-on Z(I) results for graphene on copper and freestanding graphene. The measured tunneling current for freestanding graphene is plotted versus the setpoint tunneling current and is displayed as an inset.

**Figure 2 | Experimental observation of thermal mirror buckling. a**, Constant-current, $Z(V)$ data sets on suspended graphene acquired using the labeled setpoint currents. A double arrow indicates that the scan is reversible. **b**, Measured tunneling current as a function of the tip bias for the 4.0 nA data set. **c**, Topography line profile extracted from the STM image recorded in conjunction with the 4.0 nA data set. It represents 400 s in time, with the 4.0 nA $Z(V)$ measurement being performed between the two data points which bookend the large permanent jump. **d**, Illustration of a biased STM sample near a buckled graphene membrane, which is the initially buckled graphene in our molecular dynamics simulation

**Figure 3 | Setup for the MD simulations. a**, An illustration of the model used for MD simulations. This is a circular sheet of graphene 0.18 μm wide, containing 1.1 million atoms, and broken into four regions as described in the Methods. The STM tip is above the central point. **b**, Equilibrium temperature profile of the system shown in **a** when the central region is held at 500 K and the outer region at 300 K.

**Figure 4 | MD demonstration of thermal buckling. a**, Height of the initial buckled graphene state with the bias voltage set to 3 V and the central temperature set to 300 K. **b**, Height of the buckled graphene 50 ps after the bias voltage was changed to 0.22 V and the central temperature was increased to 500 K. A mirror buckling effect due to the applied thermal load can be seen in the center. **c**, The upper (red) Gaussian-shaped curve was extracted from **a** along $x = 0$. The lower (blue) curve was extracted from **b** at $x = 0$. **d**, Magnified view of the peaks shown in **c**. The arrow indicates a 3 nm reversal in the height from a convex to concave configuration.

**Figure 5 | Simulated height-time trajectories under the STM tip. a**, Time trajectory of the central atom (solid line) and average height of the whole central region (dashed line) when $T_c = 300$ K (red, upper curves) or 500 K (green, lower curves) in the MD computations for the tip over a bump in graphene, as depicted below the graph. **b**, The same as **a** except with the tip over a depression. The height change is significantly larger because the electrostatic force and thermal load now act together to move graphene in the same direction. **c-h**, Electro-thermal-mechanical switching device with two inputs and one output. The **c-d-e-h** path shows thermal switching from input 0 to input 1, while the **c-f-g-h** path shows electrostatic switching from input 0 to input 1.

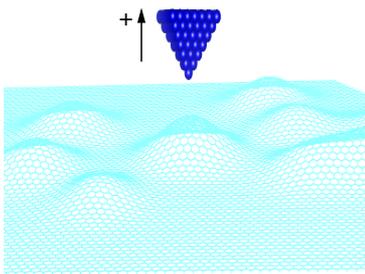 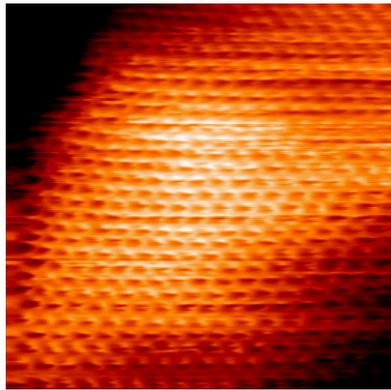 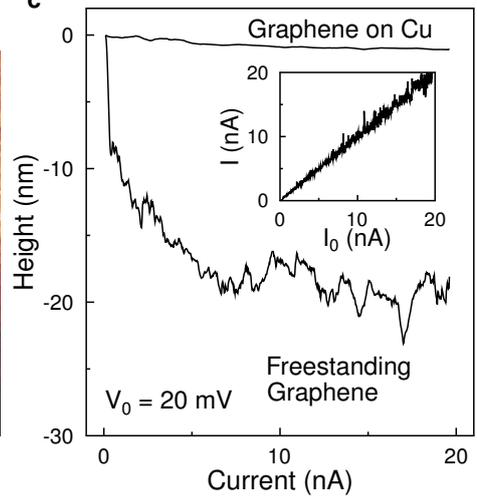

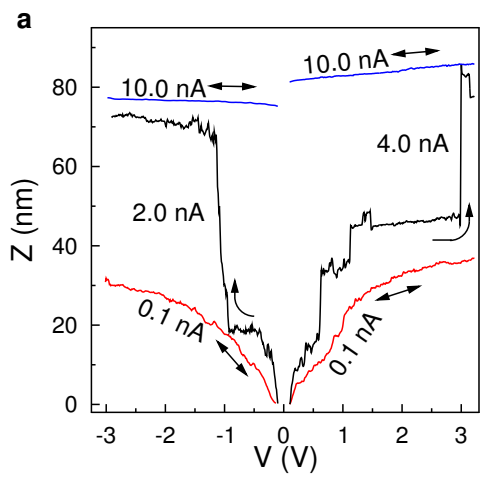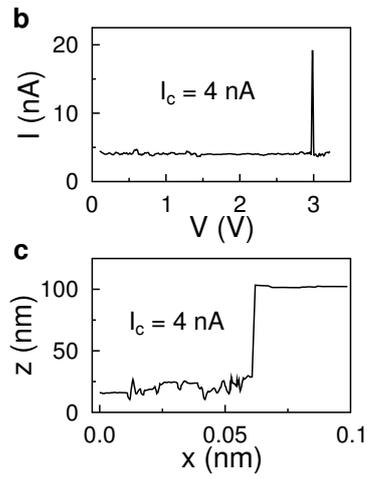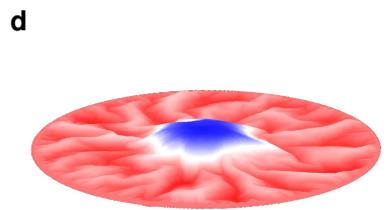

**a** 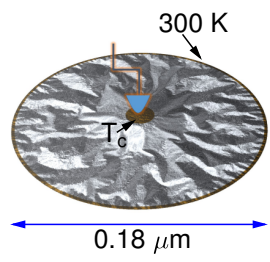

**b** 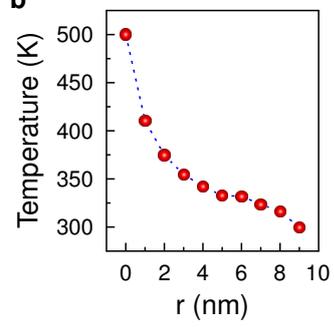

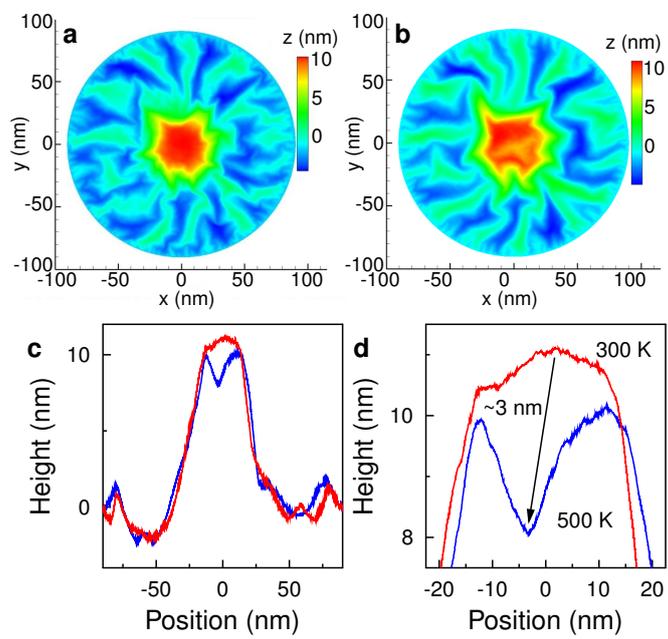

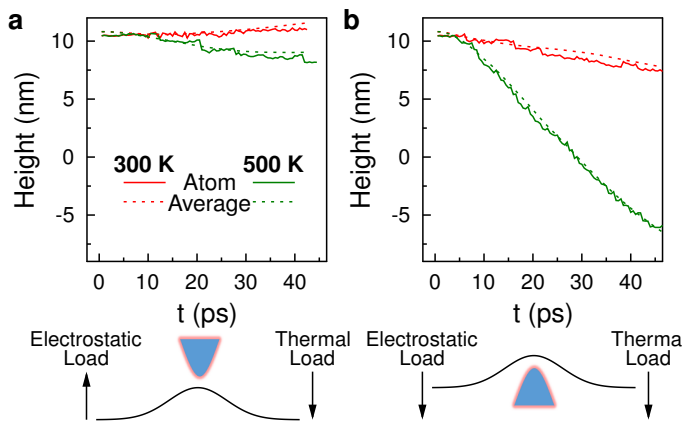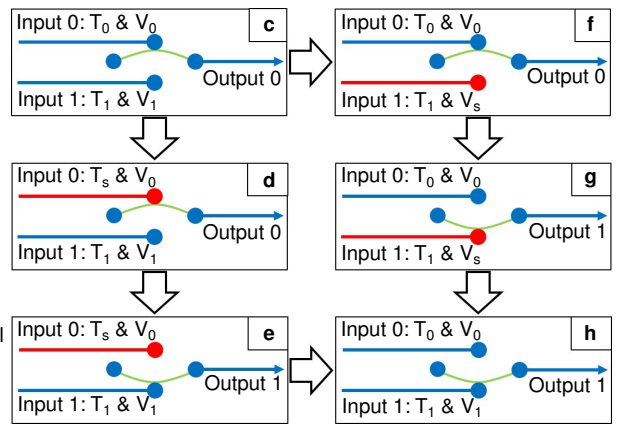